\documentclass[conference]{IEEEtran}
\IEEEoverridecommandlockouts
\usepackage{cite}
\usepackage{amsmath,amssymb,amsfonts}
\usepackage{algorithmic}
\usepackage{graphicx}
\usepackage{textcomp}
\usepackage{xcolor}

\usepackage{graphicx}
\usepackage{subfigure}
\usepackage{tikz}
\usetikzlibrary{arrows.meta, positioning}
\usepackage{diagbox}
\usepackage{url}

\def\BibTeX{{\rm B\kern-.05em{\sc i\kern-.025em b}\kern-.08em
    T\kern-.1667em\lower.7ex\hbox{E}\kern-.125emX}}
\begin{document}

\title{Converting IEC 61131-3 LD into SFC Using Large Language Model: Dataset and Testing
}

\author{\IEEEauthorblockN{Yimin Zhang}
\IEEEauthorblockA{\textit{CISTER / Faculty of Engineering}\\
\textit{University of Porto}\\
Porto, Portugal\\
0009-0005-0746-315X}
\and
\IEEEauthorblockN{Mario de Sousa}
\IEEEauthorblockA{\textit{Faculty of Engineering}\\
\textit{University of Porto}\\
Porto, Portugal\\
0000-0001-7200-1705}
}
% \and
% \and
% \and
% \and
% \and

\maketitle

\begin{abstract}
In the domain of Programmable Logic Controller (PLC) programming, converting a Ladder Diagram (LD) into a Sequential Function Chart (SFC) is an inherently challenging problem, primarily due to the lack of domain-specific knowledge and the issue of state explosion in existing algorithms.
However, the rapid development of Artificial Intelligence (AI) - especially Large Language Model (LLM) - offers a promising new approach.

Despite this potential, data-driven approaches in this field have been hindered by a lack of suitable datasets.
To address this gap, we constructed several datasets consisting of paired textual representations of SFC and LD programs that conform to the IEC 61131-3 standard.

Based on these datasets, we explored the feasibility of automating the LD-SFC conversion using LLM.
Our preliminary experiments show that a fine-tuned LLM model achieves up to 91\% accuracy on certain dataset, with the lowest observed accuracy being 79\%, suggesting that with proper training and representation, LLMs can effectively support LD-SFC conversion.
These early results highlight the viability and future potential of this approach.
\end{abstract}

\begin{IEEEkeywords}
Programmable Logic Controller, IEC 61131, Ladder Diagram, Sequential Function Chart, Large Language Model, Few-shot Learning, Fine-tuning
% component, formatting, style, styling, insert
\end{IEEEkeywords}

% Introduction
\section{Introduction} \label{sec:introduction}
Graphical programming languages for Programmable Logic Controller (PLC) programming are the preferred choice among engineers due to their closer correspondence with physical processes.
In the IEC 61131 standard~\cite{iec61131-3v3}, five programming languages are specified, among which three - Ladder Diagram (LD), Function Block Diagram (FBD), and Sequential Function Chart (SFC) - are graphical programming languages.
In some scenarios, only LD programs are provided; on the other hand, there is a substantial amount of legacy LD code that needs to be maintained and updated.
Many of these LD programs implement state-based controllers of a Discrete Event System (DES) that could be better understood by maintainers if expressed in an SFC.

Despite their widespread use, the automated generation and conversion of graphical programming languages remain longstanding challenges - particularly the automatic conversion from LD to SFC~\cite{vitormario}.
This difficulty arises primarily from two factors: first, existing algorithms lack sufficient domain knowledge, which hampers their ability to accurately describe component behaviors; second, they are susceptible to the state explosion problem, which poses significant challenges to scalability.

As a result, recent LLM-based studies have shifted focus to Structured Text (ST), a textual language in IEC 61131-3 that more similar to general-purpose programming languages, such as C or Python.
Building upon recent advances in LLMs these studies have demonstrated promising results in the automatic generation of code written in ST.
However, the performance of LLMs is highly dependent on massive volumes of training data, which are notably lacking in the field of industrial control.
In practice, participants in the industrial automation exhibit a general reluctance to make their codes publicly available~\cite{kang2025retrieval}.
Insufficient data volume may even compromise the credibility and reproducibility of the results~\cite{kilian2024universal},~\cite{liu2024agents4plc}.
Table~\ref{tab:dataset_summary} provides a summary of the datasets used in these studies and highlights their respective limitations.

\begin{table*}[!htbp]
\centering
\caption{Overview of the Datasets Employed in Related Work}
\begin{tabular}{l|l|l|l}
\hline
\textbf{Related Work} &\textbf{Dataset Volume} &\textbf{Involving Languages} &\textbf{Limitation}\\% &\textbf{Disclosure Status} &\checkmark 
\hline
% \hline
\cite{koziolek2023chatgpt} &100 &ST, FBD, SFC &Limited quantity, with only 10 examples per category.\\
\hline
\cite{koziolek2024rag} &50+ (500+ for RAG) &ST, FBD &Limited quantity.\\%585
\hline
\cite{koziolek2024image} &3 &ST &Limited quantity.\\
\hline
\cite{fakih2024llm4plc} &596 / 40 (train / test) &ST &Contains publicly available online content.\\
\hline
\cite{koziolek2024case} &10 &ST, FBD &Limited quantity.\\
\hline
\cite{kilian2024universal} &21 &ST &Limited quantity.\\
\hline
\cite{liu2024agents4plc} &23 &ST &Limited quantity.\\
\hline
\cite{haag2024training} &1300 / 200 (train / test) &ST &Contains publicly available online content.\\
\hline
\cite{yang2024multi} &914 &ST &Contains publicly available online content.\\
% \cite{yang2024multi} &718/151/45 (OSCAT/Siemens SCL/authentic) &ST &\checkmark &Contains publicly available online content.\\
\hline
\cite{kang2025retrieval} &13124 / 500 / 500 &LD &The data is not publicly available.\\
\hline
\end{tabular}
\label{tab:dataset_summary}
\end{table*}

Another key attribute we consider is the disclosure status, i.e., whether the dataset contains content that is publicly available online.
This is crucial, as the training corpora of LLMs often include a vast amount of open-access web content.
Consequently, if test sets overlap with training data, it also raises concerns regarding the validity of the evaluation~\cite{fakih2024llm4plc},~\cite{haag2024training},~\cite{yang2024multi}.

With the goal of using LLMs to automatically convert IEC 61131-3 LD programs into SFC, we did some preliminary experiments~\cite{yimin} and concluded that further teaching of LLMs was needed to reach useful results, but lacked datasets to do it with.  
We therefore begin by constructing several datasets of SFC-LD text representations, leveraging the relative ease of converting an SFC into an LD, i.e., we generate both the textual representation of SFCs and their corresponding LD equivalents (also in textual formats).
Utilizing these datasets, along with the strong text understanding capabilities of LLMs, we investigate the problem of LD-SFC conversion in textual form.

Although many LLMs are regarded as multi-modal models and capable of directly generating images, our previous experiments~\cite{yimin} showed that the output images were often overly stylized or required highly detailed prompts, which limited their practical utility.
This leads us to choose to conduct our experiments using textual representations instead.

% 值得注意的是，这些文本表示并非自然语言描述，而是标准中的定义。
% For SFC, we adopt the standardized syntax (defined in IEC 61131-3) of representing these programs in textual format since this syntax consists on textually enumerating the steps and the transitions - i.e. it does not follow a graphical format.
% For LD, we chose a different route as the textual syntax for representing LD defined in IEC 61131-3 requires that the LD programs be represented in graphical format using textual symbols to make the drawings - this would be more difficult for an LLM to parse.
% We therefore chose to represent the LD programs by expressing the logic expressions they contain as a list of textually represented logic expressions (similar to ST expressions).

The rest of the paper is structured as follows.
Section~\ref{sec:sota} reviews the related work.
Section~\ref{sec:dataset} provide a detailed description of how to construct the datasets, as well as the statistical characteristics of the datasets.
Section~\ref{sec:methodology} describe the methodology of LD-SFC conversion experiments.
Section~\ref{sec:experiments} showcases the results we got.
Section~\ref{sec:conclusion} concludes the paper.

\section{State of the Art} \label{sec:sota}
Recent studies demonstrate a strong interest among researchers in leveraging LLMs for PLC programming.
In~\cite{koziolek2024case}, Koziolek et al. attempt to generate PLC test cases for IEC 61113-3 function blocks using LLM.
The results demonstrate that this approach can efficiently produce test cases within a short time.
However, a notable limitation lies in erroneous assertions.

In~\cite{kilian2024universal}, Tran et al. primarily evaluated the capabilities of five LLMs in generating ST code.
However, the test set consists of only 21 samples, which is not very convincing.
Moreover, the study does not incorporate manual validation; instead, it relies solely on conventional metrics used in LLM evaluation.

In~\cite{liu2024agents4plc}, Liu et al. proposed a multi-agent framework to automate the generation of ST programs.
The overall task is decomposed into several specialized agents, including retrieval agent, planning agent, coding agent, debugging agent, and validation agent, thereby leveraging the multi-modal capabilities of LLMs across different sub-tasks.
However, the dataset employed in this study is notably limited, containing only 23 examples in total.
In particular, only 7 examples pertain to medium problems, which significantly undermines the generalization of the results.

In~\cite{haag2024training}, Haag et al. proposed an online feedback Direct Preference Optimization (DPO) method to generate ST programs.
It utilizes compiler outputs to construct guidance datasets online, thereby iteratively refining model parameters.
However, the test set also contains public content, and the training data is derived from Python using LLMs.
Despite these efforts, the evaluation methodology exhibits certain limitations: it relies on the LLM itself as an expert to assess the semantic correctness of the generated programs.

In~\cite{yang2024multi}, in order to overcome the limitations of available public resources, Yang et al. first developed two specialized libraries: one comprising successful case studies, including both requirements and corresponding code, and another consisting of a public instruction set tailored to ST.
Furthermore, they designed a Retrieval-Augmented Generation (RAG) mechanism to efficiently retrieve relevant cases from these libraries.
To enhance the reliability of code generation, they also implemented a self-improvement loop incorporating an integrated syntax and semantic checker customized for ST.
Feedback from the checker is subsequently analyzed by the underlying LLM to iteratively refine and optimize the generated outputs.

In~\cite{kang2025retrieval}, Kang et al. applied LLMs to LD generation,
They introduce a two-stage training strategy that combines RAG and preference learning, achieving significant improvements in LD generation performance.
In essence, they transformed the graphical generation problem into a textual representation by leveraging XML format, and then utilized the text processing capabilities of LLMs to complete the conversion.
Unfortunately, their training dataset is not publicly available, which reflects a broader challenge in the field of PLC programming and industrial control: the scarcity of datasets, the lack of open sharing, and the difficulty of accessing private industrial data owned by companies.

In summary, existing studies either leverage LLMs to process textual programming languages or to handle the textual representations of graphical programming languages.
Our work adopts a similar strategy.
Prior to these data-driven methods, rule-based methods either lacked sufficient domain knowledge~\cite{Falcione1992} or suffered from state explosion problem~\cite{VimalUnderstanding},~\cite{vitormario}.

Notably, the IEC 61131-3 standard provides a textual representation for SFC, and combined with prior research findings, SFC emerges as a suitable entry point to build up datasets and tackle the LD-SFC conversion challenge using LLMs.
To the best of our knowledge, previous research has not explored the use of LLMs for LD-SFC conversion.

% This raises concerns regarding the logical correctness and reliability of the generated results.
% Reference [2] also advocates for the construction of datasets that encompass a wide range of scenarios; however, no concrete implementation has been observed to date.
% The reported final accuracy of 39\% is relatively low compared to other processing pipelines.
% Dataset
\section{Creation and Characteristics of SFC-LD Datasets} \label{sec:dataset}
As state previously, our objective is to create pairs of SFC and LD programs that are semantically equivalent which will be used to teach LLMs on how to convert LD programs into SFC programs.
Taking advantage of the fact that automatic conversion from SFC to LD is trivial, we start by generating random SFCs, and converting each to their LD equivalents.

It must be emphasized that authentic real-world data yield the most effective training outcomes.
However, in the absence of such real-world industrial control datasets, we are compelled to construct synthetic datasets.
Given that the objective at this stage is to validate a preliminary idea, the chosen methodology is considered appropriate for exploratory purposes.
In fact, the dataset adopted in~\cite{haag2024training} also use synthetic data that derived from Python.

\subsection{Simplified Scenarios}
Based on our previous studies~\cite{yimin}, LLMs demonstrate a better understanding of SFC.
This can be attributed to the structural simplicity of SFC compared to LD.
SFCs involve fewer fundamental elements - namely: steps, transitions, branches, and actions.
To reduce complexity and focus on the core structure, we exclude actions in our current setup.
As a result, the SFCs in our dataset comprise only steps, transitions, and branches.
Meanwhile, the input and output variables are disregarded, serving as a starting point for this preliminary study.
More complex scenarios will be explored in future research.

\subsection{Construction Methodology}
\begin{figure}[!ht]
  \centering
  \subfigure[Sequential Structure]{
    \includegraphics[width=0.4\linewidth]{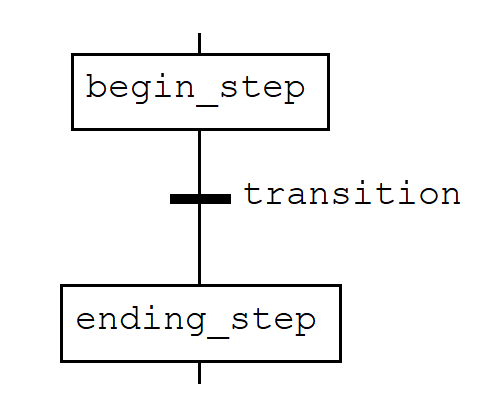}
    \label{fig:sub1}
  }
  \hfill
  \subfigure[Simultaneous Branch Structure]{
    \includegraphics[width=0.6\linewidth]{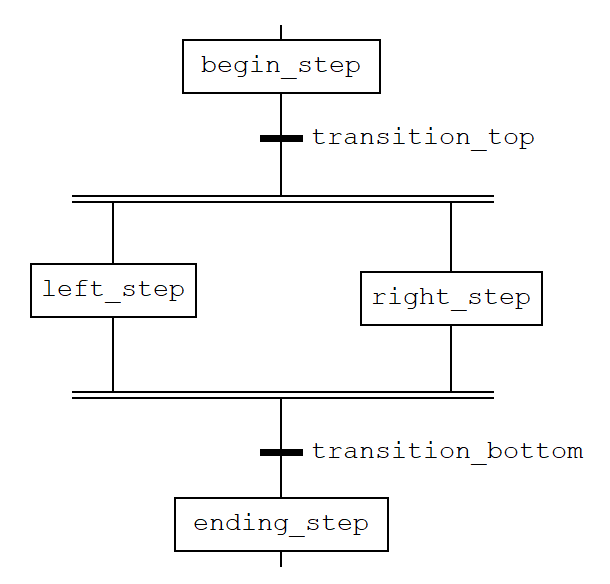}
    \label{fig:sub2}
  }
  \hfill
  \subfigure[Selective Branch Structure]{
    \includegraphics[width=0.95\linewidth]{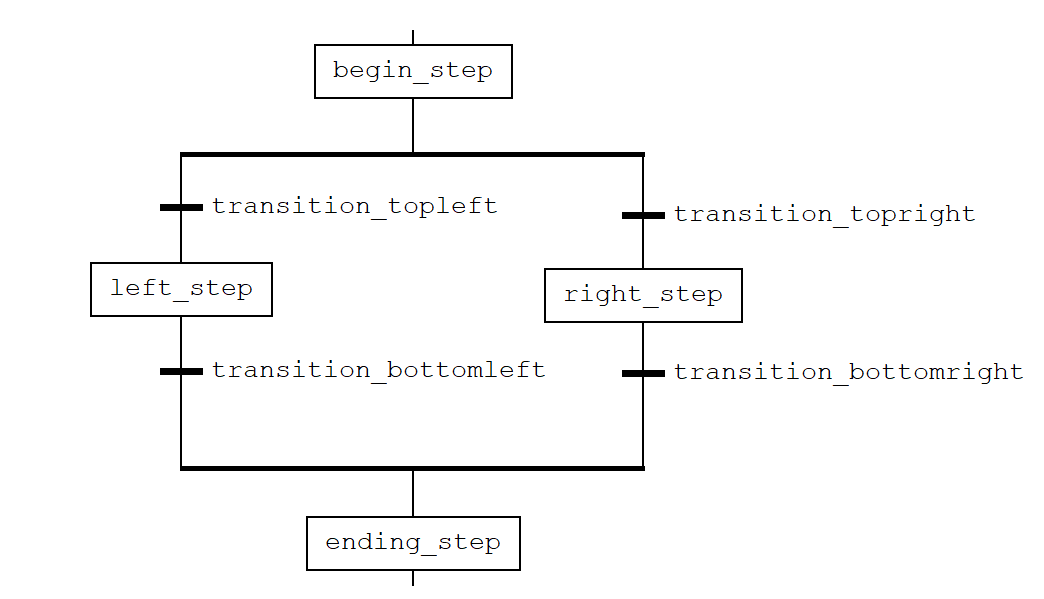}
    \label{fig:sub3}
  }
  \caption{Three Basic Structures.}
  \label{fig:3basic}
\end{figure}

We define three fundamental SFC structures: sequential structure, simultaneous branch structure, and selective branch structure, as illustrated in Figure~\ref{fig:3basic}.
Each structure begins with a designated begin step and terminates at an end step.

These structures are recursively generated with predefined probabilities.
In line with practical considerations, sequential structures tend to occur more frequently, while simultaneous and selective branches are less common.
Accordingly, we assign a relatively higher probability $p_{seq}$ to sequential structures and lower probabilities to the other two types ($p_{sim}$, $p_{sel}$).

\begin{figure}[!ht]
    \centering
    \includegraphics[width=0.8\linewidth]{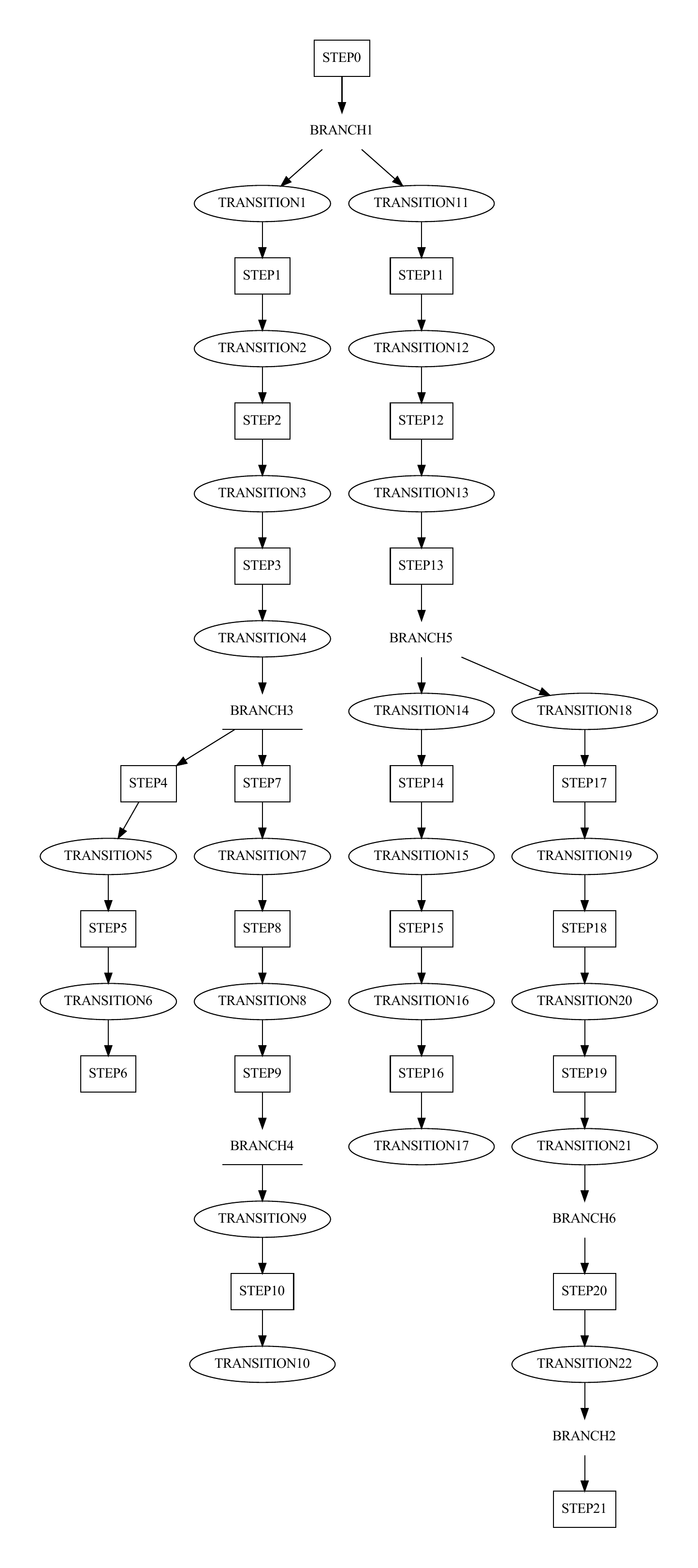}
        \put(-170,136){\color{black}\vector(1,-0.3){35}}  %画一个箭头
        \put(-80,115){\color{black}\vector(1,-0.25){40}}
        \put(-127,52){\color{black}\vector(1,-0.1){88}}
        % \put(25,63){\color{red}\textbf{This is a feature}}  % 添加注释
    \caption{An Example from Dataset 2 of The Generated Data Structures.}
    \label{fig:1example}
\end{figure}

To simplify the generation process, we specify that the begin and end steps of both the simultaneous and selective branch structures do not participate in recursion.
Recursion is applied only to the left and right branches within these structures.
Starting from the three structural patterns, the entire SFC is generated recursively as controlled by another parameter, depth ($d$).
An example of the generated data structures from Dataset 2 is illustrated in Figure~\ref{fig:1example} that exhibits considerable complexity.

\subsection{Parameter Descriptions}
% Branch numbers???
By adjusting the probability parameter $(p_{seq}, p_{sim}, p_{sel})$ and the recursion depth $d$, we can generate SFCs of arbitrary complexity.
If the objective is to increase the occurrence of branch structures, their associated probabilities must be raised accordingly.
However, when the recursive depth is set too high, it leads to structural explosion, resulting in excessively large and complex programs.
Conversely, when the probabilities for branch structures are low, larger depths can be tolerated without significantly increasing the overall complexity.
We experimented with a variety of parameter configurations and finally create four dataset.
These datasets will be made publicly available on GitHub.\footnote{\url{https://github.com/yimin-up/Converting_IEC_61131_LD_into_SFC_Using_LLM_Dataset_and_Testing.git}}
Table~\ref{tab:parameter} summarizes the parameter settings for the four datasets.

\begin{table}
    \centering
    \caption{Parameters for Each Dataset}
    \begin{tabular}{c|c|c|c|c}
    \hline
        \diagbox{Parameter}{Dataset} &1  &2  &3  &4\\
        \hline
        Sequential structure probability &0.5  &0.8  &0.9  &0.9\\
        Simultaneous branch probability &0.3  &0.1  &0.1  &0\\
        Selective branch probability &0.2  &0.1  &0  &0.1\\
        Recursion depth &3  &6  &6  &6\\
        Number of examples &120$^{\mathrm{a}}$  &100  &100  &100\\
        \hline
        \multicolumn{5}{l}{$^{\mathrm{a}}$100 for training, 20 for validation.}
    \end{tabular}
    \label{tab:parameter}
\end{table}

\subsection{Dataset Statistics}
\begin{figure}[!ht]
  \centering
  \subfigure[Boxplot of the Number of Steps]{
    \includegraphics[width=0.95\linewidth]{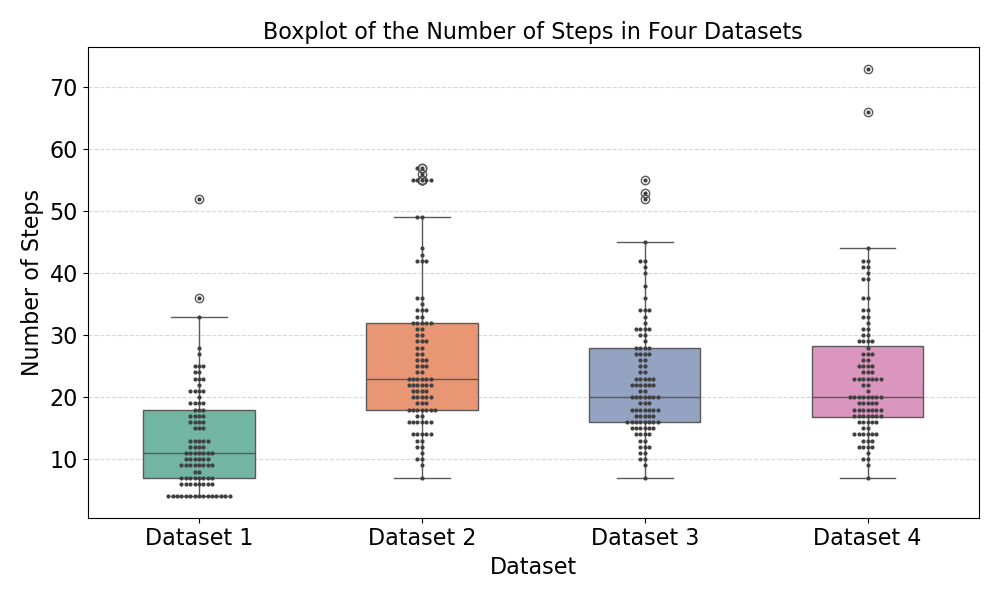}
    \label{fig:boxplot_step}
  }
  % \hfill
  \subfigure[Boxplot of the Number of Transitions]{
    \includegraphics[width=0.95\linewidth]{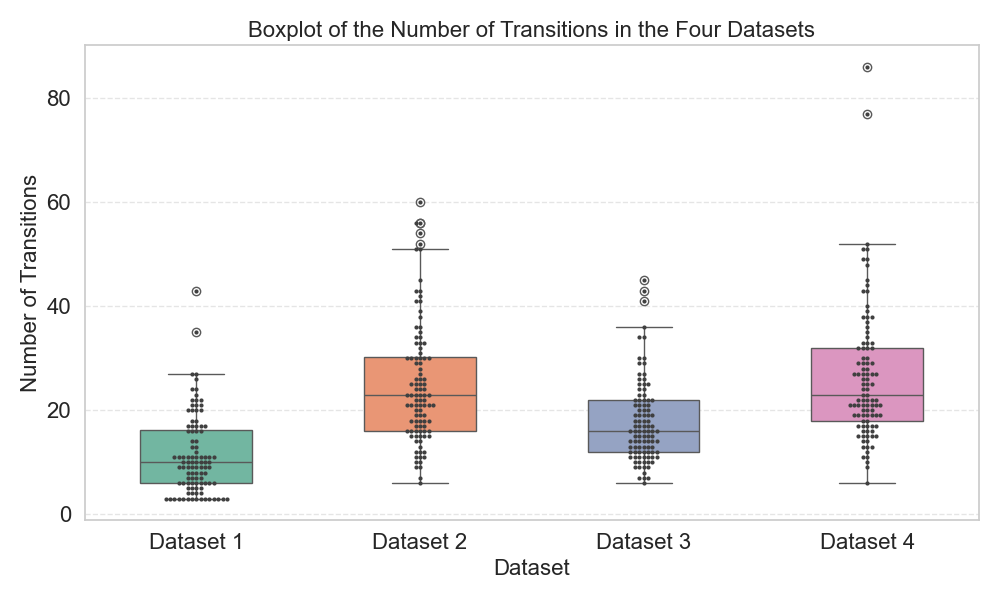}
    \label{fig:boxplot_transition}
  }
  \caption{Boxplot of the Number of Steps and Transitions}
  \label{fig:boxplot}
\end{figure}

We collected statistics on the number of steps and number of transitions across the four datasets, Figure~\ref{fig:boxplot_step} $\sim$ Figure~\ref{fig:boxplot_transition}, which serve as indicators of program complexity.

% \begin{figure}[!t]
%   \centering
%   \subfigure[Normal Distribution Approximation of Number of Steps]{
%     \includegraphics[width=0.8\linewidth]{picture/kde_step.png}
%     \label{fig:kde_step}
%   }
%   % \hfill
%   \subfigure[Normal Distribution Approximation of Number of Transitions]{
%     \includegraphics[width=0.8\linewidth]{picture/kde_transition.png}
%     \label{fig:kde_transition}
%   }
%   \caption{Normal Distribution Approximation of Number of Steps and Transitions}
%   \label{fig:kde}
% \end{figure}

\subsection{SFC-LD Conversion}
\begin{figure}[!ht]
  \centering
  \subfigure[Example SFC]{
    \includegraphics[width=0.25\linewidth]{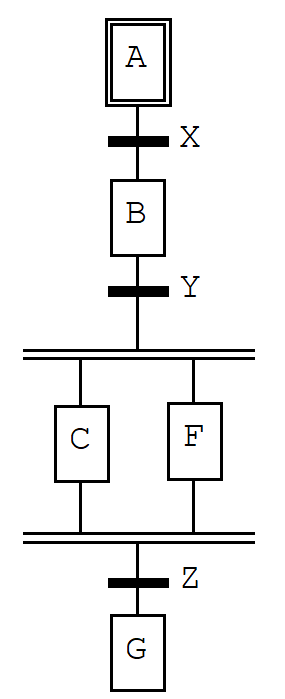}
    \label{fig:example_SFC}
  }
  \hfill
  \subfigure[Equivalent LD]{
    \includegraphics[width=0.5\linewidth]{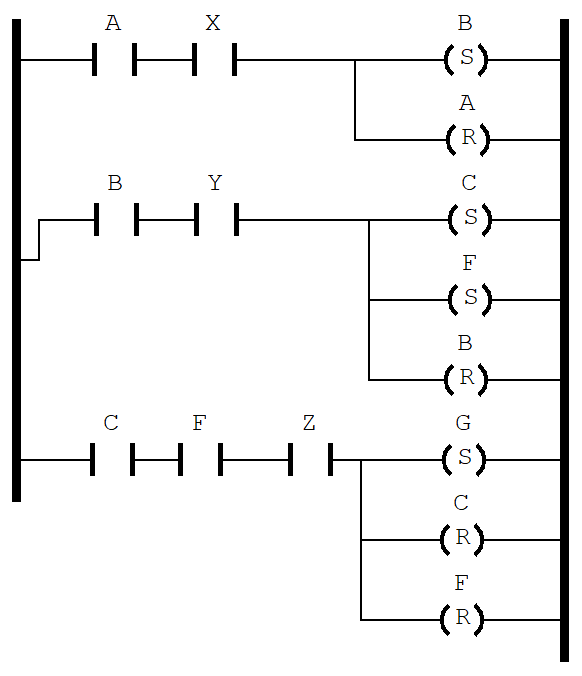}
    \label{fig:example_LD}
  }
  \caption{Example SFC and Its Equivalent LD.}
  \label{fig:example}
\end{figure}

As the inverse problem of LD-SFC conversion, SFC-LD conversion is easier and can be achieved through many methods.
A straightforward method involves the use of SET/RESET coil.
Figure~\ref{fig:example} shows an example SFC and its equivalent LD which further demonstrates the methodology for deriving an equivalent LD representation from the underlying data structure of an SFC, highlighting the feasibility of reverse transformation.
% 这进一步阐释了如何从SFC的数据结构获得同等LD表达。

The IEC 61131-3 standard specifies a formal textual representation for SFC.
However the standard does not define a formal textual representation for LD.
Nevertheless, compilers internally convert LD into intermediate expressions.
We adopt equivalent expressions as the textual format of LD in our study.
The expressions represent the LD in Figure~\ref{fig:example_LD} are:
\begin{center}
    \begin{verbatim}
    IF A AND X:
        B := 1;
        A := 0;
    IF B AND Y:
        C := 1;
        F := 1;
        B := 0;
    IF C AND F AND Z:
        G := 1;
        C := 0;
        F := 0;
    \end{verbatim}
\end{center}

\section{Experiments Methodology} \label{sec:methodology}

\subsection{Overview}
\begin{figure}[!ht]
    \centering
    \includegraphics[width=0.75\linewidth]{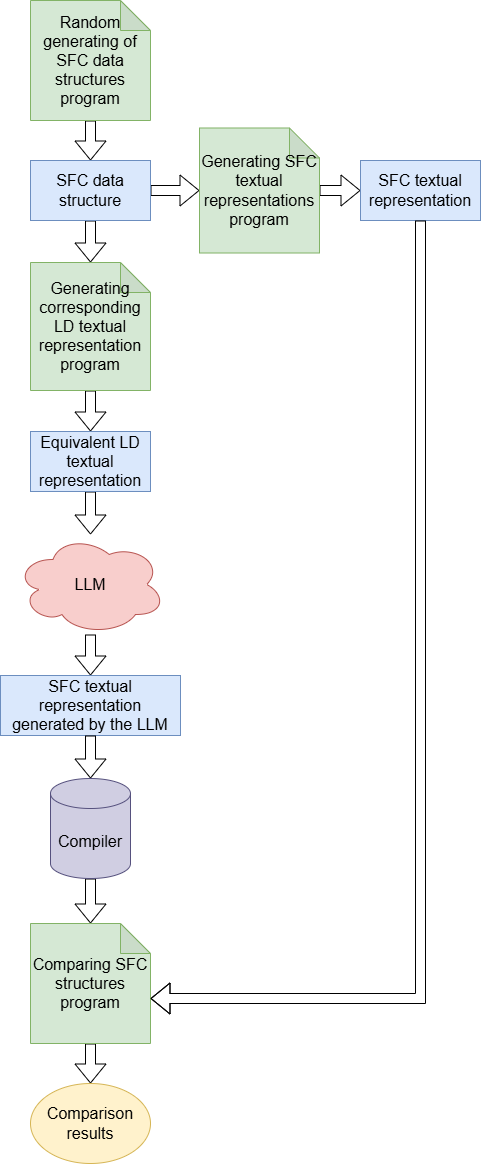}
    \caption{LD to SFC Conversion Experiments Overview.}
    \label{fig:flow}
\end{figure}

Our process begins with the generation of SFC data structures, as described in Figure~\ref{fig:flow}.
For each structure, we obtain both the textual representation of the SFC and the equivalent LD textual representation.
We input the LD text into the LLM, explicitly instructing it that this is a LD textual description, and request it to generate the equivalent SFC.
For convenience, we denote the SFC generated by the LLM as SFC (LLM).

The output produced by the LLM is first compiled using a open-source compiler MatIEC~\cite{MATIEC} to check for syntactic correctness.
The next step is data structure check, which involves structural comparison between the SFC (LLM) and the original ground truth SFC.
This objective is accomplished through performing reverse engineering to recover their internal data structures from SFC (LLM).
This enables us to perform automated structural comparison to determine equivalence.

\subsection{LLM Model}
As of the time of writing, three GPT-4 models are publicly available: ``gpt-4o-2024-08-06", ``gpt-4o-mini-2024-07-18", and ``gpt-4-0613".
Among these, ``gpt-4o-mini" is officially recommended for most use cases due to its favorable trade-off between cost and performance.
Given that this work represents a preliminary investigation, we primarily employed ``gpt-4o-mini" in consideration of computational cost.
% 我们会在未来工作中考虑其他模型，例如deepseek，gemini，但在这里我们觉得没必要。fall outside the scope of this preliminary investigation and , further evaluations and comparisons can be conducted more reliably.
While we acknowledge the potential of other models such as ``Gemini"\cite{geminiteam2023gemini} and ``DeepSeek"\cite{liu2024deepseek}, we chose not to include them in the current study, as we believe such comparisons are better suited for future work when a more comprehensive dataset is established.

We explored several approaches to generating SFC text representations from LD, including zero-shot learning, few-shot learning, and model fine-tuning.
Due to the limited performance of zero-shot learning, which consistently failed to produce compilable programs, we excluded it from further evaluation.

\subsection{Experiments}
We conducted experiments on Datasets 2, 3, and 4.
To maximize the diversity of results, each dataset contains 100 unique samples.
Dataset 2 contains all three basic structures.
Dataset 3 is composed exclusively of sequential and simultaneous branch structures, while Dataset 4 includes sequential and selective branch structures only.
Dataset 1, used for fine-tuning, features a balanced structural composition.
As illustrated in Figures~\ref{fig:boxplot} of Section~\ref{sec:dataset}, Dataset 1 shows clear statistical differences from the other datasets, which helps mitigate the risk of overfitting to any specific pattern.

\section{Results and Discussion} \label{sec:experiments}
We define three metrics: syntax check pass rate, structural check pass rate, and joint pass rate, which respectively represent:
\begin{itemize}
    \item The proportion of SFC (LLM) passing MatIEC's syntax check during compilation.
    \item The proportion of SFC (LLM) passing structural comparison check.
    \item The proportion of SFC (LLM) passing both syntax check and structural check.
\end{itemize}

\subsection{Few-shot learning vs fine-tuning} \label{exp:finetune}
\begin{figure}
    \centering
    \includegraphics[width=0.9\linewidth]{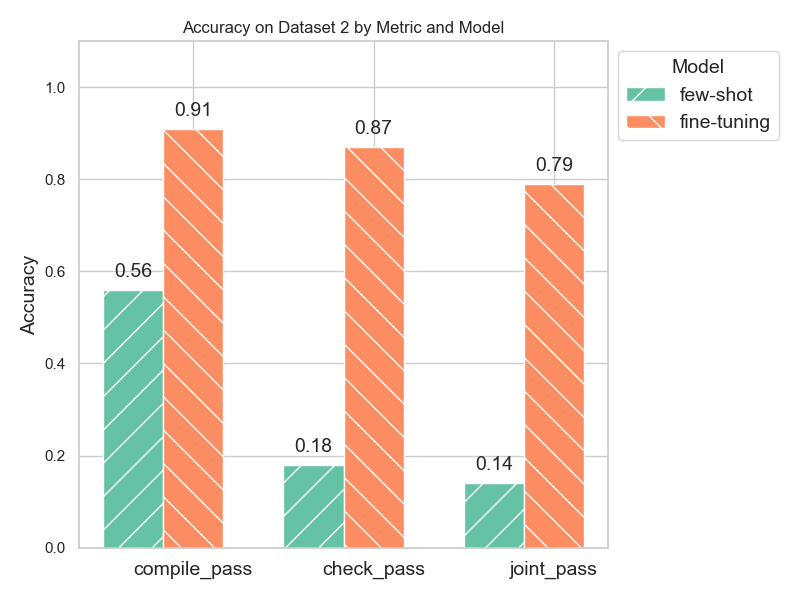}
    \caption{Accuracy on Dataset 2}
    \label{fig:d2_single}
\end{figure}

\begin{figure}
    \centering
    \includegraphics[width=0.9\linewidth]{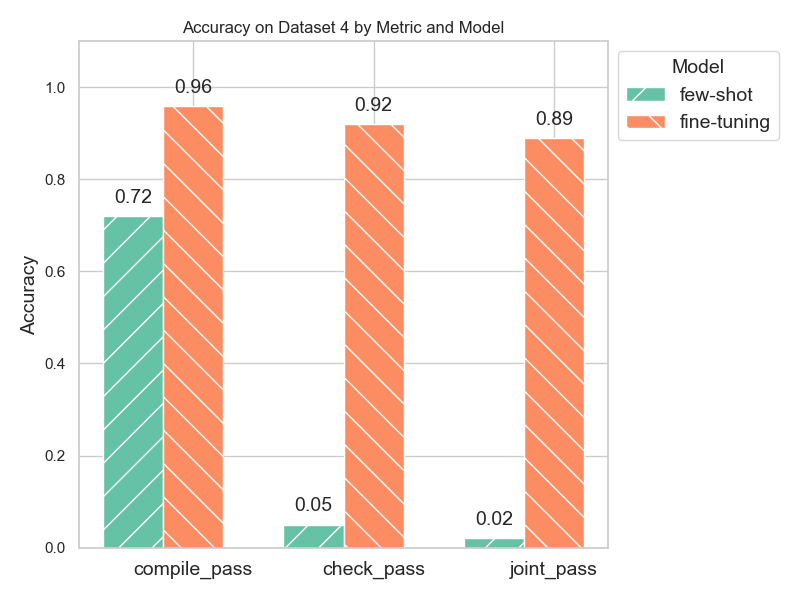}
    \caption{Accuracy on Dataset 4 (selective branches only)}
    \label{fig:d4_single}
\end{figure}

Figures~\ref{fig:d2_single} $\sim$ \ref{fig:d4_single} present the LD-SFC conversion pass rates on Dataset 2 $\sim$ Dataset 4.

\begin{figure}
    \centering
    \includegraphics[width=0.9\linewidth]{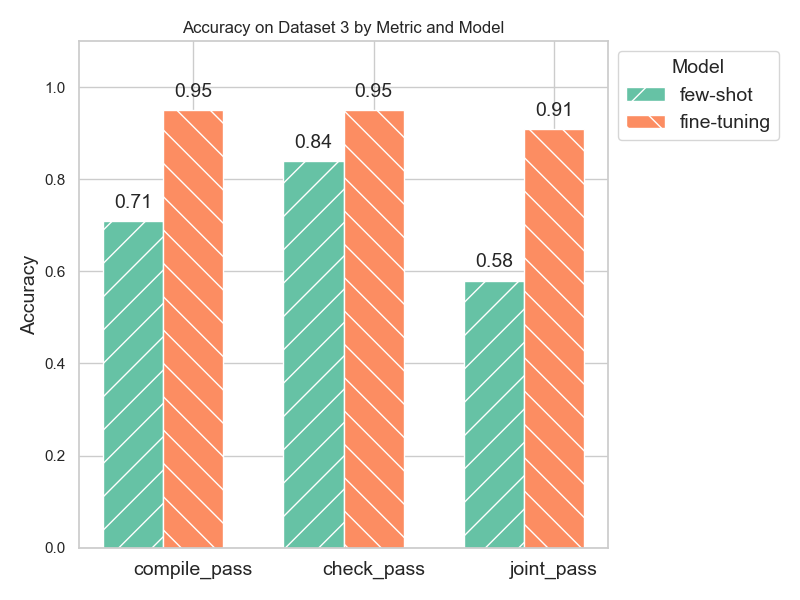}
    \caption{Accuracy on Dataset 3 (simultaneous branches only)}
    \label{fig:d3_single}
\end{figure}

Experiments on Dataset~2 clearly demonstrate the significant advantage of the fine-tuned model.
The joint pass rate reaches 79\% for the fine-tuned model, while the few-shot learning approach only achieved 14\%.

To investigate the reasons behind these failures, we analyzed the error patterns and identified three common causes:
\begin{itemize}
    \item Typos.
    \item Missing variable declaration.
    \item Omission of one or more branches in simultaneous branch structures.
\end{itemize}

The first two are relatively easy for LLMs to fix.
To address the third issue specifically, we designed Dataset~3 and Dataset~4, where $p_{sim}$ or $p_{sel}$ is set to 0.
Surprisingly, the pass rate on Dataset~3 was significantly higher than on Dataset~4, which contradicts the trends observed on Dataset~2.
This inconsistency need further investigation.

\subsection{The Impact of Program Complexity}
\begin{figure*}[!htbp]
    \centering
    \includegraphics[width=0.8\linewidth]{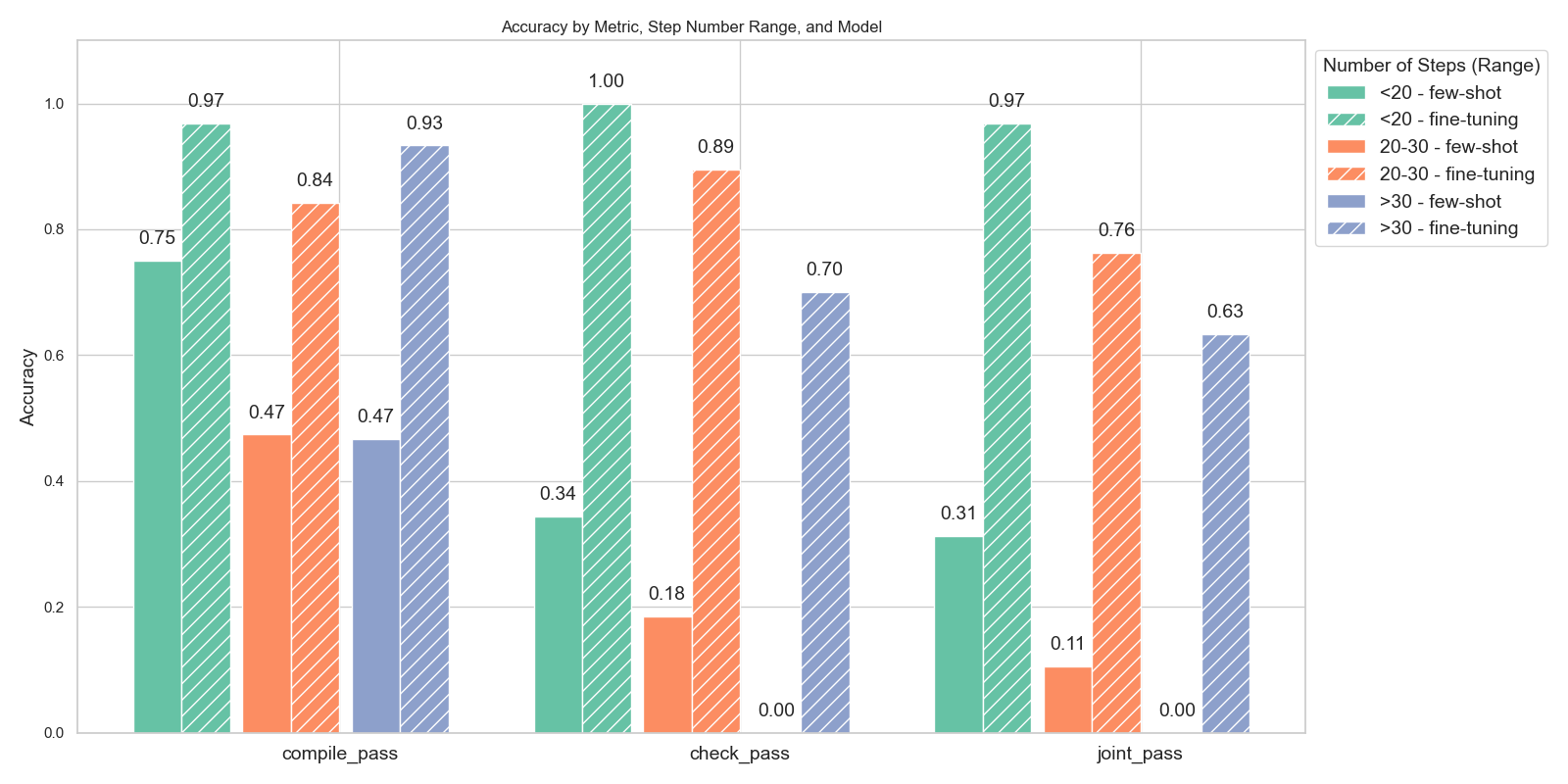}
    \caption{Accuracy by Step Number Range, and Model on Dataset 2}
    \label{fig:step_num2}
\end{figure*}

\begin{figure*}[!htbp]
    \centering
    \includegraphics[width=0.8\linewidth]{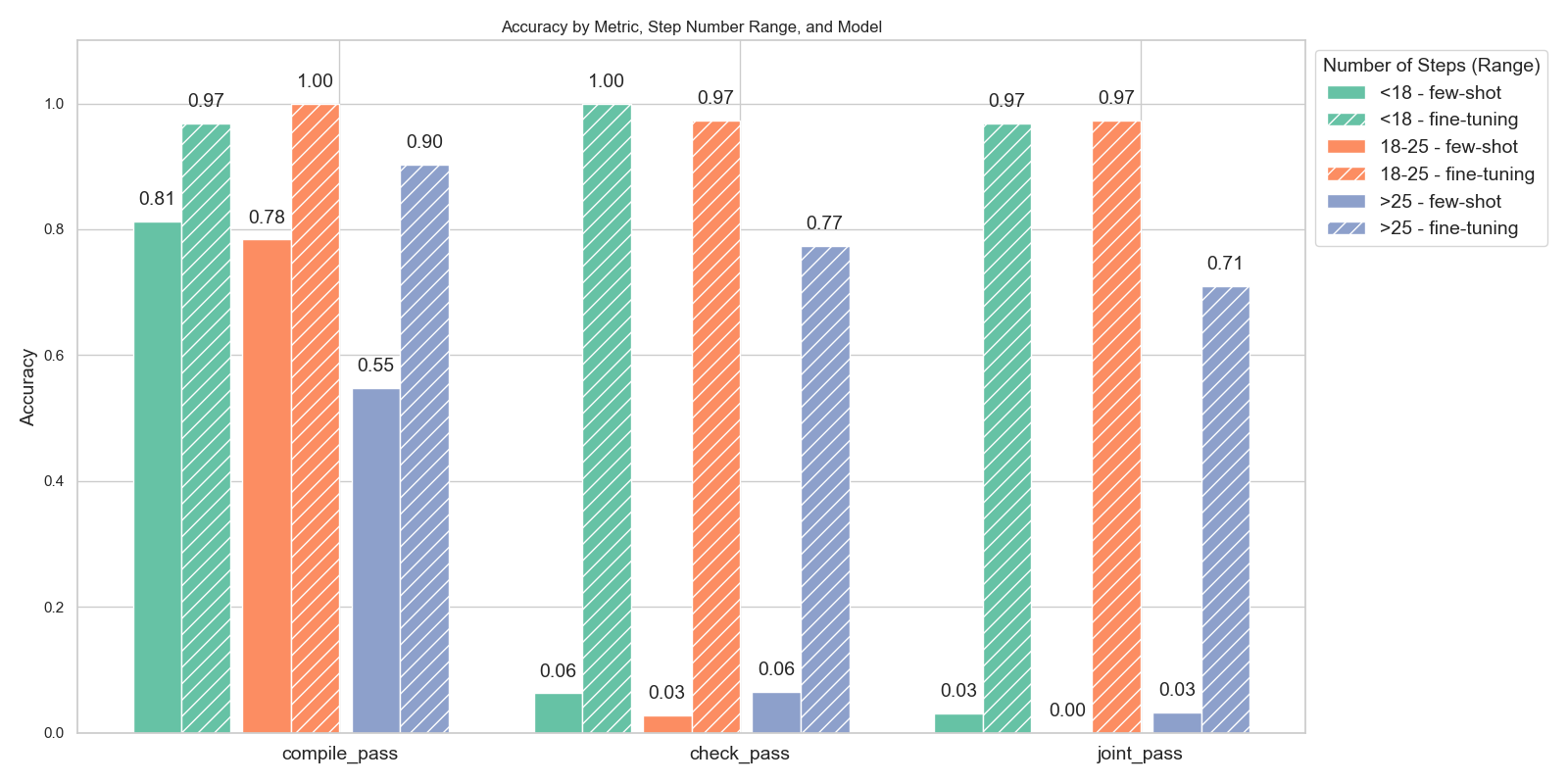}
    \caption{Accuracy by Step Number Range, and Model on Dataset 4}
    \label{fig:step_num4}
\end{figure*}

In general, the more complex a program is, the more difficult it becomes to convert, and thus, the lower the expected pass rate.
To verify this assumption, we divided each dataset into three groups based on the number of steps in each program, and evaluated the pass rate within each group individually.
The results are shown in Figures~\ref{fig:step_num2}~$\sim$~\ref{fig:step_num4}.

\begin{figure*}[!htbp]
    \centering
    \includegraphics[width=0.8\linewidth]{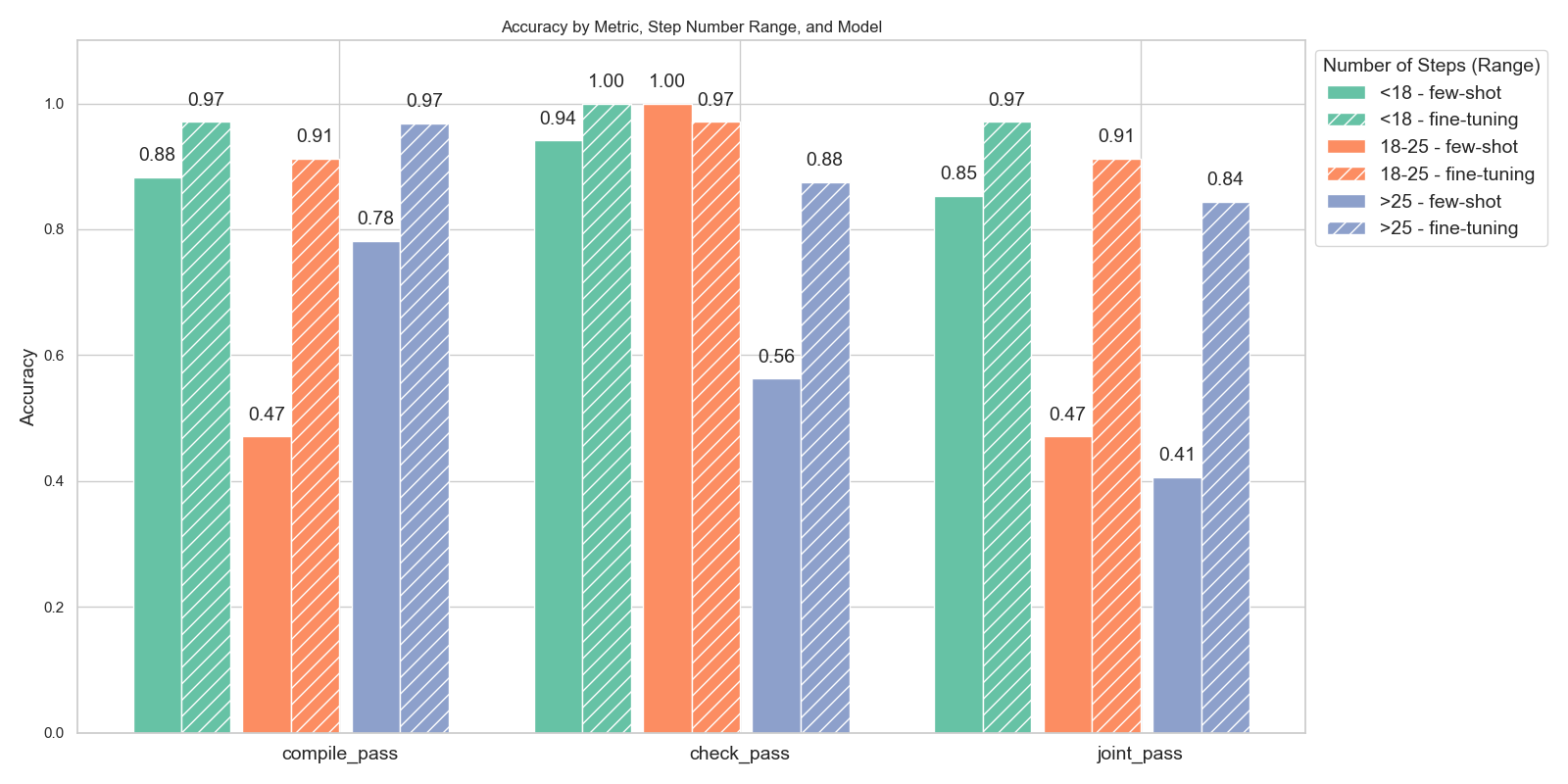}
    \caption{Accuracy by Step Number Range, and Model on Dataset 3}
    \label{fig:step_num3}
\end{figure*}

Based on the dataset distribution described in Section~\ref{sec:dataset}, we carefully designed the grouping strategy to ensure that each group contains approximately 1/3 of the total number of programs.
This was done to maintain statistical balance and avoid having too few samples in any single group, which would weaken the reliability of the conclusions.
The detailed grouping information is presented in Table~\ref{tab:number_group}.
\begin{table}[!htbp]
    \caption{Grouping Based on the Number of Steps for Dataset 2-4}
    \centering
    \begin{tabular}{c|p{1.5cm}|p{1.5cm}|p{1.5cm}}
        \hline
        \textbf{Dataset} &\multicolumn{3}{c}{\textbf{Grouping Based on the Number of Steps}} \\
        \cline{2-4} &{Group 1} &{Group 2} &{Group 3} \\ \hline
        2 &$<$20  &20-30  &$>$30 \\
        3 &$<$18  &18-25  &$>$25 \\
        4 &$<$18  &18-25  &$>$25 \\
        \hline
    \end{tabular}
    \label{tab:number_group}
\end{table}

For both structural check and joint check, our experiments show a clear trend: as program complexity increases, the accuracy decreases.
However, in the case of syntax checking, this trend is not observed - the accuracy does not consistently degrade with increased complexity.
This phenomenon may arise because syntax checking is susceptible to various non-semantic interferences, such as the typos mentioned previously.

\subsection{Discussion}
As an initial exploration, our study primarily aims to point out a feasible solution: by converting graphical LD and SFC representations into textual formats, we can better leverage the strengths of LLMs in processing text.

It is important to acknowledge that the feasibility in our experiments is, in part, due to our simplification strategy in generating LD representations.
These simplifications, while suitable for experimentation, do not yet reflect the full complexity of industrial control programs.

It is also necessary to clarify that, at this initial stage, we intentionally limit the amount of data exposed to the LLM, considering that some models may retain user-provided inputs.
We did not adopt larger-scale datasets - such as generating 1,000 or 10,000 samples - nor did we explore more sophisticated agent architectures or complex augment strategies e.g. RAG or multi-agent strategy.

\section{Conclusion and Future Work} \label{sec:conclusion}
\subsection{Conclusion}
Our research explores the feasibility of using LLM to convert LD into SFC in textual formats.
Due to the scarcity of existing dataset, we constructed a SFC-LD dataset, which, to the best of our knowledge, is the first dataset of its kind.
Experimental results show that, even without applying additional augment techniques, a fine-tuned ``gpt-4o-mini" model can achieve an accuracy of 79\%.
Specifically, the fine-tuned model achieves up to 91\% accuracy on certain dataset without selective branches.
This provides a new perspective for addressing the LD-SFC conversion problem.

\subsection{Future work}
One major limitation of our work lies in the gap between generated datasets and real-world industrial programs.
While our datasets enabled controlled experimentation, it cannot fully capture the complexity, diversity of actual industrial applications.
Expanding the dataset to better simulate real-world scenarios is a crucial next step.
% 例如一个微小但重要的改进是：随机命名系统是必要的。现在的命名系统遵循了非常简单的规则，但实际中程序员命名可能非常随意。
% It appears that the generation of test SFCs follows a very simple naming convention. It is likely that charts written by humans are much more messy. The fine tuning to the simple naming might skew results.
For example, a minor but important improvement is the introduction of a randomized naming system. The current naming scheme follows overly simplistic rules, whereas in practice, programmers often use highly arbitrary naming conventions.

Another key extension is to support additional textual formats commonly used in industrial automation.
Standards such as PLCopen XML provide structured, machine-readable representations of control logic, which are well-suited to LLM-based workflows.
Since these formats are text-based, they align naturally with the strengths of current models.

% To summarize, future work include:
% \begin{itemize}
%     \item Extend datasets to reflect real-world diversity.
%     \item Extend datasets to XML format or other standard format.
% \end{itemize}

% \subsection{Limitation}

% An ideal future direction would involve collecting and annotating real industrial examples of LD–SFC pairs. Unfortunately, such examples are scarce and often proprietary, making this direction both essential and challenging. Community efforts to curate open, diverse, and domain-relevant datasets could be a game-changer.

% Several papers have begun exploring this direction using proprietary data; however, due to the closed nature of their datasets, reproducibility and insight into the specific modeling challenges remain limited.

\bibliographystyle{IEEEtran}
\bibliography{myrefs.bib}

\end{document}